\documentclass[aps,prd,superscriptaddress,twoside,twocolumn,nofootinbib,10pt,%
showpacs,floatfix]{revtex4-1}

\usepackage{amsmath,amssymb}
\usepackage{graphicx,bm}
\usepackage{slashed}
\usepackage{dcolumn}
\usepackage{epstopdf}
\usepackage{ulem} 
\usepackage[usenames]{color}

\allowdisplaybreaks

\renewcommand\sout{\bgroup \color{red} \ULdepth=-.5ex \ULset}

\begin{document}

\title{Topological dark matter from the theory of composite electroweak symmetry breaking}

\author{Yong-Liang Ma}
\email{yongliangma@jlu.edu.cn}
\affiliation{Center for Theoretical Physics and College of Physics, Jilin University, Changchun,
130012, China}

\date{\today}
\begin{abstract}
The lightest electroweak baryon as a topological object is investigated by using a general effective Lagrangian of composite electroweak symmetry breaking and the spin-independent electroweak baryon-nucleon scattering cross section is calculated. We explicitly show the masses of the electroweak baryons and the cross section as functions of the Peskin-Takeuchi $S$ parameter and the ratio of the masses of axial-vector and vector composite bosons. We find that it is acceptable to regard the electroweak baryon as a dark matter candidate and the even number of technicolor is favored.
\end{abstract}


\maketitle


\section{Introduction}

\label{sec:intro}

The mechanism of electroweak symmetry breaking (EWSB) is still an open question in particle physics (for a recent review, see, e.g., Ref.~\cite{Peskin:2012vu}) even after the confirmation of the Higgs boson~\cite{:2012gk,:2012gu}. Among the models of new physics, the composite EWSB mechanism is the interesting one (see~\cite{Hill:2002ap} and references therein).

The basic character of the composite EWSB mechanism is the following: Given the existence of a fundamental non-Abelian gauge theory, such as technicolor (TC) theory, above the scale of EWSB, one explains the Higgs boson in standard model (SM) of particle physics as a composite particle, similar to the light scalar meson in QCD. It was found that, to have a phenomenologically acceptable, TC theory should have a softly scale dependent (nearly conformal) gauge coupling constant~\cite{Yamawaki:1985zg}, that is, a scale-invariant/walking/conformal technicolor (SWC-TC) theory. And, near the conformal window, the vector and axial-vector composite states are almost degenerate~\cite{Appelquist:1998xf} and, the Peskin-Takeuchi $S$ parameter should be reduced~\cite{Appelquist:1991is,Sundrum:1991rf,Appelquist:1998xf}.

In the composite EWSB mechanism, one can immediately expect that there are composite EW baryons, like what happens in QCD, in addition to the composite bosons. In the literature, the spectrum of the TC baryon as a soliton has been studied more than 30 years~\cite{Gipson:1980ag,Gipson:1983cm,D'Hoker:1984jq,Nussinov:1985xr} and it was found that the mass of the lightest one is a few TeV. Recently, The topological TC baryon was discussed by several groups. In Refs.~\cite{Murayama:2009nj,Joseph:2009bq,Gillioz:2010mr,Gillioz:2011dj}, in composite Higgs models, the TC baryon as a skyrmion was discussed. Based the LHC constraints, bounds of the TC baryon mass was analysed~\cite{Ellis:2012bz,Ellis:2012cs,Kitano:2016ooc} and, in the framework of standard model Higgs Lagrangian, the upper bounds of the lightest EW baryon could be $\sim 38$~GeV~\cite{Matsuzaki:2016iyq}. Since the topological EW/TC baryon carries conserved topological charge, it is a stable objects. In addition, by a suitable choice of the charges of its constituents, the EW baryon can be a charge neutral object. These features indicate that the EW baryon could be a dark matter (DM) candidate. Combing the data and experimental measurements, the possible EW/TC baryon contribution to DM was discussed in the literature~\cite{Murayama:2009nj,Joseph:2009bq,Gillioz:2010mr,Gillioz:2011dj,Bagnasco:1993st,Ellis:2012bz,Ellis:2012cs,Kitano:2016ooc,Matsuzaki:2016iyq}.

Since the EWSB mechanism is based on the strongly coupled gauge theory, the chiral symmetry of the fundamental theory, here TC, can be broken due to the condensate of the techni-fermion at certain scale, similar to the chiral symmetry breaking in QCD. The purpose of the present work is to investigate the composite EW baryon using the skyrmion approach~\cite{Skyrme:1961vq} which has been widely used in particle and condense matter physics (see Refs.~\cite{Ma:2016gdd,BReditor} for recent reviews and references therein). We use an effective theory of strongly coupled theory with nonlinear realized chiral symmetry including the vector boson techni-$\rho$ and axial-vector boson techni-$a_1$ based on the generalized hidden local symmetry (GHLS)~\cite{Bando:1987ym,Bando:1987br} so that the effect of the mass difference between techni-$\rho$ and techni-$a_1$ on the EW baryon can be studied. After integrating the techni-$\rho$ and techni-$a_1$, one can express the Skyrme parameter $e$ in terms of parameter $R$ which is the ratio of the masses of techni-$a_1$ and techni-$\rho$ and Peskin-Takeuchi $S$ therefore investigate the $R$ and $S$ dependence of the EW baryon properties for the purpose to show the EW baryon properties near the conformal window. By using such calculated EW baryon properties and with respect to the observation constraint, we calculate the elastic scattering between the EW baryon without EW quantum numbers and nuclei.

We find that, near conformal window, a smaller $S$ parameter leads to a lighter bosonic EW baryon with mass around a few hundred GeV and this small mass could be further reduced to a few ten GeV when the masses of techni-$\rho$ and techni-$a_1$ are nearly degenerate. However, near the conformal window, the mass of fermionic EW baryon is larger than 100 TeV, so that its freeze-out temperature is much larger than the EW scale. We consider the elastic scattering between the bosonic EW baryon and nuclei and find that the cross section is below the most stringent constraint for the weakly interacting massive (WIMP) DM from PandaX-II~\cite{Tan:2016zwf} so that it is reasonable to regard the bosonic EW baryon as a DM candidate. However, for the fermionic EW baryon, the saturation of the relic density require the underlying TC theory far away from conformal window. So that, with respect to the mass spectrum, the odd number of technicolor is not favored.

This paper is organized as follows: In Sec.~\ref{sec:ghls} we generally explore the lightest EW baryon properties using an effective Lagrangian including techni-$\pi$, techni-$\rho$ and techni-$a_1$. In Sec.~\ref{sec:dark} we discuss the possibility to regard the EW baryon as a DM candidate. Our remarks and discussions are given in Sec.~\ref{sec:dis}.

\section{The lightest EW baryon from general hidden local symmetry}

\label{sec:ghls}

We investigate the lightest EW baryon spectrum with respect to the typical features of SWC-TC by using a chiral effective Lagrangian of strongly coupled gauge theory at low energy scale, the generalized hidden local symmetry (GHLS)~\cite{Bando:1987ym,Bando:1987br}, including techni-$\pi$, techni-$\rho$ and techni-$a_1$. Here, we only consider the Lagrangian with the minimal number of derivatives (for a complete next to leading order Lagrangian, see~\cite{Ma:2004sc}). Following Refs.~\cite{Bando:1987ym,Bando:1987br}, in GHLS for $N_{TF}$ techni-flavors, we decompose the techni-$\pi$ field $U(x)$ as
\begin{eqnarray}
U(x) & = & \xi_L^\dag(x)\xi_M(x)\xi_R(x).
\end{eqnarray}
Under chiral transformation, the relevant quantities transform as
\begin{eqnarray}
\xi_{L,R}(x) & \to & \hat{g}_{L,R}(x)\xi_{L,R}^\prime(x)g_{L,R}^\dag ; \nonumber\\
\xi_M(x) & \to & \xi_M^\prime = \hat{g}_L(x) \xi_M(x)\hat{g}_R^\dagger(x).
\end{eqnarray}
where $g_{L,R} \in [SU(N_{\rm TF})_{L,R}]_{\rm global}$ are the elements of the chiral symmetry and $\hat{g}_{L,R}(x) \in [SU(N_{\rm TF})_{L,R}]_{\rm local}$ stand for the elements of GHLS. Introducing the gauge fields $L_\mu(x)(R_\mu(x))$ corresponding to the local symmetry $\hat{g}_{L,R}(x)$, the effective Lagrangian with the minimal number of derivatives could be written as
\begin{eqnarray}
{\cal L} & = & a {\cal L}_V + b {\cal L}_A + c {\cal L}_M + d {\cal L}_\pi + {\cal L}_{\rm kin} \label{eq:ghls}
\end{eqnarray}
with $a, b, c$, and $d$ being dimensionless constants and
\begin{widetext}
\begin{eqnarray}
{\cal L}_V & = & f_\pi^2{\rm Tr}\left[\frac{D_\mu \xi_L\cdot \xi_L^\dag + \xi_MD_\mu \xi_R \cdot \xi_R^\dag \xi_M^\dag}{2i}\right]^2 , \quad {\cal L}_A = f_\pi^2{\rm Tr}\left[\frac{D_\mu \xi_L\cdot \xi_L^\dag - \xi_MD_\mu \xi_R \cdot \xi_R^\dag \xi_M^\dag}{2i}\right]^2 \, , \nonumber\\
{\cal L}_M & = & f_\pi^2{\rm Tr}\left[\frac{D_\mu \xi_M \cdot \xi_M^\dag }{2i}\right]^2 , \quad {\cal L}_\pi = f_\pi^2{\rm Tr}\left[\frac{D_\mu \xi_L\cdot \xi_L^\dag - \xi_MD_\mu \xi_R \cdot \xi_R^\dag \xi_M^\dag - D_\mu \xi_M \cdot \xi_M^\dag}{2i}\right]^2 \, ,\nonumber\\
{\cal L}_{\rm kin} & = & {} - \frac{1}{4 g^2}F^{(L)}_{\mu\nu}F^{(L)\mu\nu} - \frac{1}{4 g^2}F^{(R)}_{\mu\nu}F^{(R)\mu\nu} \, ,
\end{eqnarray}
\end{widetext}
where $f_\pi = v_{\rm EM}$ is the EW scale. The covariant derivatives and the field strengths in above are defined as
\begin{eqnarray}
D_\mu \xi_{L} & = & \partial_\mu \xi_{L} - i L_\mu \xi_L , \nonumber\\
D_\mu \xi_{R} & = & \partial_\mu \xi_{R} - i R_\mu \xi_R ,  \nonumber\\
D_\mu \xi_{M} & = & \partial_\mu \xi_{M} - i L_\mu \xi_L + \xi_M R_\mu , \nonumber\\
F_{\mu\nu}^{(L)} & = & \partial_\mu L_\nu - \partial_\nu L_\mu - i \left[L_\mu, L_\nu\right] , \nonumber\\ F_{\mu\nu}^{(R)} & = & \partial_\mu R_\nu - \partial_\nu R_\mu - i \left[R_\mu, R_\nu\right] . \label{eq:defghls}
\end{eqnarray}
In GHLS, the local symmetries $[SU(N_{\rm TF})_{L,R}]_{\rm local}$ are broken so that both vector field $
V_\mu = \frac{1}{2}\left(R_\mu + L_\mu\right)$ and axial-vector field $A_\mu = \frac{1}{2}\left(R_\mu - L_\mu\right)$ acquire masses. After the local symmetry breaking the Lagrangian in the unitary gauge is
\begin{eqnarray}
{\cal L} & = & \left(b + d\right)f_\pi^2 {\rm Tr}\left[\hat{\alpha}_{\perp \mu}\hat{\alpha}_{\perp}^{\mu}\right] + a f_\pi^2  {\rm Tr}\left[\hat{\alpha}_{\parallel \mu}\hat{\alpha}_{\parallel}^{\mu}\right] \nonumber\\
& &{} + 2 b {\rm Tr}\left[A_{\mu}\hat{\alpha}_{\perp}^{\mu}\right] + \left(b + c\right)f_\pi^2 {\rm Tr}\left[A_{\mu}A^{\mu}\right]  \nonumber\\
& & {} - \frac{1}{2 g^2}{\rm Tr}\left[F^{(V)}_{\mu\nu}F^{(V)\mu\nu}\right] - \frac{1}{2 g^2}\left[F^{(A)}_{\mu\nu}F^{(A)\mu\nu}\right] \, , \label{eq:ghlsfix}
\end{eqnarray}
where
\begin{eqnarray}
\hat{\alpha}_{\parallel\mu}^{} & = & \frac{1}{2i} \left( D_\mu \xi_R^{}\cdot
\xi_R^\dagger + D_\mu \xi_L^{}\cdot \xi_L^\dagger \right), \nonumber\\
\hat{\alpha}_{\perp\mu}^{} & = & \frac{1}{2i} \left( D_\mu \xi_R^{}\cdot
\xi_R^\dagger - D_\mu \xi_L^{}\cdot \xi_L^\dagger \right),\nonumber\\
F^{(V)}_{\mu\nu} & = & \partial_\mu V_\nu - \partial_\nu V_\mu - i \left[V_\mu, V_\nu\right] - i \left[A_\mu, A_\nu\right] \, , \nonumber\\
F^{(A)}_{\mu\nu} & = & \partial_\mu A_\nu - \partial_\nu A_\mu - i \left[A_\mu, V_\nu\right] - i \left[V_\mu, A_\nu\right] \, .
\end{eqnarray}
The covariant derivatives are defined as $ D_\mu \xi_{R,L}^{} = (\partial_\mu - i V_\mu) \xi_{R,L}^{} $ with
$\xi_L^\dagger = \xi_R^{} \equiv \xi = e^{i\pi/(2f_\pi)}$.

The techni-$\rho$ and techni-$a_1$ and their flavor partners are introduced in GHLS through substitutions $V_\mu = \frac{g}{2} \rho_{ \mu}$ and $A_\mu = \frac{g}{2} a_{1 \mu}$. From Eq.~(\ref{eq:ghlsfix}) one sees that both vector and axial-vector mesons become massive
\begin{eqnarray}
m_{a_1}^2 & = & {} \left(b + c\right) g^2 f_\pi^2 \; ,  \qquad m_\rho^2 = a g^2 f_\pi^2 . \label{eq:ghlsmass}
\end{eqnarray}
In addition, one can incorporate the external gauge fields $\mathcal{L}_\mu$ and $\mathcal{R}_\mu$ by gauge the global symmetry $[SU(N_{\rm TF})]_{\rm global}$ through the following substitutions in Eq.~(\ref{eq:defghls}):
\begin{eqnarray}
D_\mu \xi_{L} & = & \partial_\mu \xi_{L} - i L_\mu \xi_L + i \xi_L \mathcal{L}_\mu ,\nonumber\\
D_\mu \xi_{R} & = & \partial_\mu \xi_{R} - i R_\mu \xi_R + i \xi_R \mathcal{R}_\mu .
\end{eqnarray}
With this expressions, the $\rho$-$\gamma$ and $a_1$-$\gamma$ transition strengths $g_\rho$ and $g_{a_1}$ are obtained as~\cite{Bando:1987ym,Bando:1987br}
\begin{eqnarray}
g_\rho & = & a f_\pi ^2 g \, , \qquad g_{a_1} = b f_\pi^2 g .\label{eq:ghlsmixing}
\end{eqnarray}

We next integrate the techni-$a_1$ and techni-$\rho$ fields through their equations of motion in consequence. After an appropriate normalization, one reduces the Lagrangian (\ref{eq:ghlsfix}) to the standard Skyrme model
\begin{eqnarray}
{\cal L} & = & f_\pi^2 {\rm Tr}\left[\alpha_{\perp \mu} \alpha_{\perp}^\mu\right]+ \frac{1}{2 e^2} {\rm Tr} \left\{\left[\alpha_{\perp \mu}, \alpha_{\perp \nu}\right]\left[\alpha_{\perp}^{\mu}, \alpha_{\perp}^\nu\right]\right\} . \label{eq:skyrmeL}
\end{eqnarray}
with $e$ as the Skyrme parameter and $f_\pi$ proportional to the EW scale $v_{\rm EM}$. In terms of the parameters in Lagrangian (\ref{eq:ghlsfix}), the Skyrme parameter is expressed as
\begin{eqnarray}
\frac{1}{2e^2} & = & \frac{1}{2g^2}\left[\left(1 - \frac{b}{b + c}\right)\left(1 + \frac{b}{b + c}\right)\right]^2 .
\end{eqnarray}
Using the expression for the $S$ parameter in terms of vector and axial-vector parameters~\cite{Peskin:1990zt,Peskin:1991sw} and Eqs.~(\ref{eq:ghlsmass}) and (\ref{eq:ghlsmixing}), one can obtain
\begin{eqnarray}
S & = & 4 \pi N_D \frac{1}{g^2}\left[1 - \left(\frac{b}{b + c}\right)^2\right] ,
\end{eqnarray}
where $N_D$ is a number of weak doublet techni-fermion which relates the EW scale and $f_\pi$ through $v_{\rm EW} = f_\pi \sqrt{N_D}$. Therefore
\begin{eqnarray}
\frac{1}{2e^2} & = & \frac{1}{2}\frac{S}{4 \pi N_D}\left[1 - \left(\frac{g_{a_1}}{g_\rho}\frac{m_\rho^2}{m_{a_1}^2}\right)^2\right] .
\end{eqnarray}
With respect to the definitions of the decay constants of the $a_1$ and $\rho$ mesons in GHLS
\begin{eqnarray}
f_{a_1}^2 & = & \left(\frac{g_{a_1}}{m_{a_1}}\right)^2, \quad f_{\rho}^2 = \left(\frac{g_{\rho}}{m_{\rho}}\right)^2 ,
\end{eqnarray}
the Skyrme parameter is rewritten as
\begin{eqnarray}
\frac{1}{2e^2} & = & \frac{1}{2}\frac{S}{4 \pi N_D}\left[1 - \frac{f_{a_1}^2m_{a_1}^2}{f_\rho^2m_\rho^2}\left(\frac{m_\rho^2}{m_{a_1}^2}\right)^2\right] \, .
\end{eqnarray}
By using  Weinberg's second sum rule
\begin{eqnarray}
f_{a_1}^2 m_{a_1}^2 & = & f_{\rho}^2 m_{\rho}^2 \, , \label{eq:winbergsr}
\end{eqnarray}
we reduce this expression to
\begin{eqnarray}
\frac{1}{e^2} & = & \frac{S}{4\pi N_D}\left(1 - \frac{1}{R^4}\right) \, ,
\label{eq:ESR}
\end{eqnarray}
where $ R = m_{a_1}/m_\rho $ which, in QCD, has the value $\sqrt{2}$. Eq.~\eqref{eq:ESR} shows that near the conformal window, i.e., $S \to 0$ and $R \to 1$, the Skyrme parameter $e$ becomes a large quantity.

The standard calculation in Skyrme model (\ref{eq:skyrmeL}) yields the mass of techni-baryon as~\cite{Adkins:1983ya}
\begin{eqnarray}
M_{\rm baryon} & = & M_{\rm sol} + \frac{j(j+1)}{2\mathcal{I}},\label{eq:massratio}
\end{eqnarray}
where the soliton mass $M_{\rm sol}$ and momentum of inertia $\mathcal{I}$ calculated as
\begin{eqnarray}
M_{\rm sol} & = & 74.1 \times \frac{f_\pi}{e}, \quad \mathcal{I} = 48.7 \times \frac{1}{e^3f_\pi} .
\end{eqnarray}
These expressions clearly show that the magnitude of the Skyrme parameter $e$ which is affected by $S$ and $R$ determines the EW baryon mass because $f_\pi$ is at the order of EWSB scale. In addition, Eq.~\eqref{eq:massratio} shows that when TC number is even, the EW baryon is a boson with the smallest mass $M_{\rm baryon} = M_{\rm sol}$ while when TC number is odd, the EW baryon is a fermion with the smallest mass $M_{\rm baryon} = M_{\rm sol} + 3/(8\mathcal{I})$.

\begin{figure}[htbp]
\begin{center}
\includegraphics[scale=0.6]{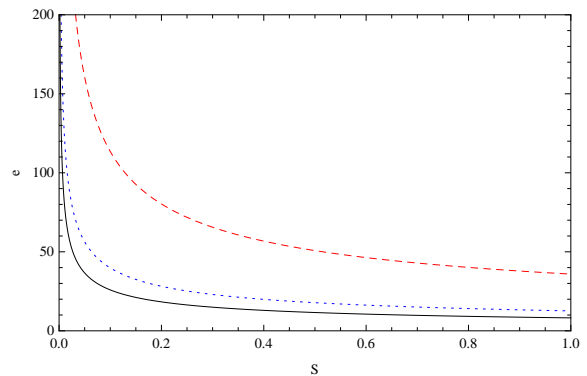}
\end{center}
\caption[]
{ The $S$ parameter dependence of the Skyrme parameter $e$ with $R = \sqrt{2}$ (solid line), $ 1.1$ (dotted line) and $1.01$ (dashed line). }
\label{fig:Sdepend}
\end{figure}

We first consider the effect of the $S$ parameter on the Skyrme parameter $e$. Here, the region of $S$ parameter is taken as $0 \le S \le 1.0$ with respect to the present fit~\cite{Olive:2016xmw}. We plot the $S$ dependence of Skyrme parameter $e$ with typical values $R = \sqrt{2}, 1.1$ and $1.01$ in Fig.~\ref{fig:Sdepend}. This figure shows that the Skyrme parameter is very sensitive to the $S$ parameter, especially when it is small, like in the region $S \leq 0.14$ at 95\% CL ~\cite{Olive:2016xmw}.

\begin{figure}[htbp]
\begin{center}
\includegraphics[scale=0.6]{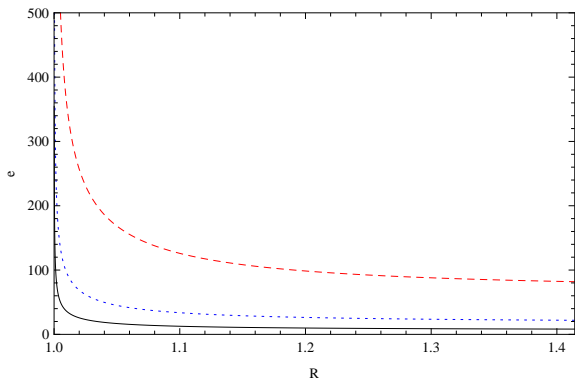}
\end{center}
\caption[]
{ The $R$ parameter dependence of the Skyrme parameter $e$ with $S = 1.0$ (solid line), $0.14$ (dotted line) and $S = 0.01$ (dashed line). }
\label{fig:Rdepend}
\end{figure}

We next turn to the effect of parameter $R$ on Skyrme parameter $e$. Here, the typical values of $S$ are taken as $S = 0.01, 0.14$ and $1.0$. We plot the $R$ dependence of the Skyrme parameter $e$ in Fig.~\ref{fig:Rdepend} by taking $1.0 < R \le \sqrt{2}$ with the upper value being the QCD case. Fig.~\ref{fig:Rdepend} shows that the smaller $R$ value, the larger Skyrme parameter $e$, i.e., the Skyrme parameter is further enhanced by the techni-$\rho$ and techni-$a_1$ degeneracy.

\begin{figure}[htbp]
\begin{center}
\includegraphics[scale=0.6]{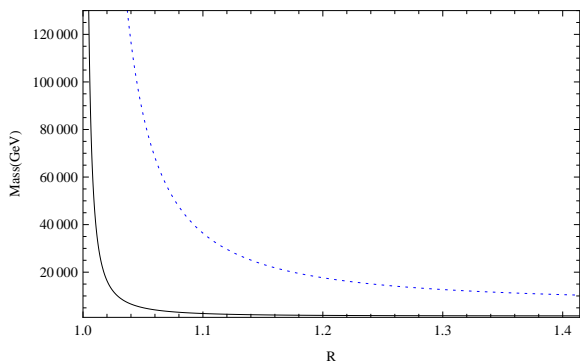}
\end{center}
\caption[]
{ The $R$ parameter dependence of the lightest bosonic EW baryon mass with $S = 1.0$ (solid line), $0.14$ (dotted line) and $0.01$ (dashed line). }
\label{fig:massghls}
\end{figure}

After a general analysing the effects of the $S$ and $R$ on the Skyrme parameter $e$. We show in Fig.~\ref{fig:massghls} the parameter $R$ dependence of the lightest bosonic EW baryon mass with $S = 1.0, 0.14$ and $0.01$ and $v_{\rm EM} = 246~$GeV. From this figure we conclude that a small $S$ or small $R$ leads to a small EW mass around a few hundred GeV.

By solving the homogeneous Bethe-Salpeter equation in the large $N_f$ QCD with the improved ladder approximation in the Landau gauge, it was found that approaching the chiral phase transition point from the broken phase, the vector and axial-vector meson masses are degenerate~\cite{Harada:2003dc}. This character should be taken into account in our present case since we are working on the conformal TC and chiral symmetry restoration is a necessary condition for the conformal invariance. The same conclusion was obtained by using GHLS if the chiral symmetry restoration arises from the Ginzberg-Landau type fixed point~\cite{Harada:2005br}. As a result, in SWC-TC, $R \simeq 1$. With respect to the situation that,  at 95\% CL, $S \leq 0.14$ at 95\% CL ~\cite{Olive:2016xmw}, the bosonic EW baryon mass should be smaller than 300 GeV. However, if we increase the value of the $S$ parameter to $1.0$, the bosonic EW baryon mass could be around 1000 GeV.

\begin{figure}[htbp]
\begin{center}
\includegraphics[scale=0.9]{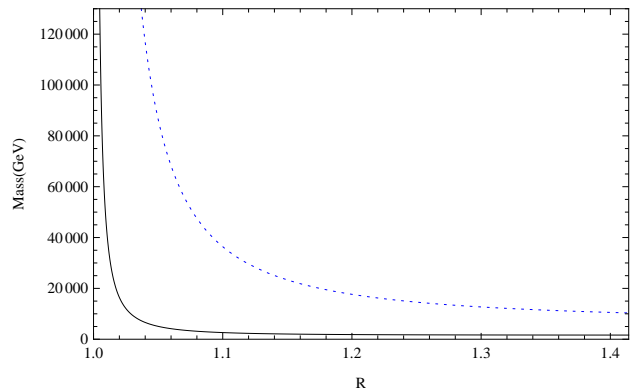}
\end{center}
\caption[]
{ The $R$ parameter dependence of the lightest EW fermionic baryon mass with $S = 1.0$ (solid line) and $0.14$ (dotted line). The plot of the mass with $0.01$ is beyond the region of the plot. }
\label{fig:massfermion}
\end{figure}

We plot in Fig.~\ref{fig:massfermion} the lightest fermionic EW baryon mass as a function of parameter $R$ with typical values of $S$. In this case, due to the contribution from collective rotation, the parameter dependence of the fermionic baryon mass is opposite to that of the bosonic baryon mass. From this result, one concludes that, for the TC theory near the conformal window, the EW baryon should be heavier than 100 TeV. Such a heavy EW baryon seems unnatural since its freeze-out temperature is much higher than the EW scale.

\section{Dark electroweak baryon}

\label{sec:dark}

The charge neutral, massive and stable topological EW baryon has attracted lots of interests in regarding it as a DM candidate. We next discuss this possibility in the present approach. With respect to the relation between the relic density of baryonic matter $\Omega_b$ and the pair annihilation cross section $\sigma_A$
\begin{eqnarray}
\Omega_bh^2 & \simeq & \frac{3 \times 10^{-27}cm^3/sec}{\langle \sigma_A v_{\rm rel}\rangle},
\end{eqnarray}
one can estimate the upper bound of Skyrme parameter as $e \simeq 150$~\cite{Kitano:2016ooc} by using $\Omega_bh^2 \simeq 0.02$ for baryonic dark matter~\cite{Olive:2016xmw} and the geometric cross section ($\sim \pi R^2$ with $R = 1/(e f_\pi)$~\footnote{We will see later that for the fermionic EW baryon this estimation is not valid.}). With this constraint, the lower limit of the bosonic EW-baryon mass is $60~$GeV while the upper limit of the fermionic mass is around $3000~$TeV which are very week constraints.

We next turn to the TC-baryon-nuclei scattering. The naive dimension counting shows that, for a bosonic TC-baryon, the charge radius give the dominant contribution while for a fermionic TC-baryon the magnetic moment is essential~\cite{Bagnasco:1993st}. Since in the skyrmion approach, the charge radius can be explicitly calculated, we study the cross section of the bosonic TC-baryon since the fermionic TC-baryon is not so natural in the present calculation.

For a charge neutral bosonic TC-baryon, it can be expressed as a complex boson field since it carries a conserved quantum number -- technibaryon number -- of a continuous symmetry. The effective interaction in terms of the TC-baryon field $\phi$ is written as
\begin{eqnarray}
{\cal L}_{\phi} & = &{} -i e g_\phi (\phi^\ast \partial_\nu\phi - \partial_\nu\phi^\ast  \phi)  \partial_\mu F^{\mu\nu},
\end{eqnarray}
with $F^{\mu\nu}$ being the field tensor of photon. Since in scalar QED $-i e (\phi^\ast \partial_\nu\phi - \partial_\nu\phi^\ast  \phi)$ is the charge density current, one can conclude that $g_\phi$ is the charge radius of charge neutral TC-baryon which, in the Skyrme model, is expressed as
\begin{eqnarray}
\sqrt{g_\phi} & = &{} 0.96 \times \frac{1}{e f_\pi}.
\end{eqnarray}
Then, in the nonrelativistic limit, the cross section of dark matter-nuclei scattering is expressed as~\cite{Bagnasco:1993st}
\begin{eqnarray}
\sigma_{\mathcal N}& = & \int_0^{E_R^{\rm max}}dE_R \frac{64\pi (Z\alpha)^2 m_N^2 g_\phi^2}{ E_R^{\rm max}(1+m_N/M_{\rm baryon})^2}|F_C(E_R)|^2.\nonumber\\
\end{eqnarray}
where $m_N$ is the nucleon mass, $F_C$ is the nuclear form factor which is normally taken the form $F_C(Q) = \exp(-Q/2Q_0)$ where $Q_0=1.5/(m_N R_0^2)$ with $R_0=(0.3 + 0.89 A^{1/3})$~\cite{Bertone:2004pz}. $E_R^{\rm max}$ is the maximum value of the recoil energy $E_R$ with magnitude $O(10)$~KeV which is taken $100$~KeV~\cite{Tan:2016zwf} in the present calculation.

Normally, the experimental data are give in terms of the normalized DM-nucleon scattering cross section $\sigma_n^{\rm SI}$ in the spin independent case. $\sigma_n^{\rm SI}$ relates to the DM-nuclei scattering cross section $\sigma_{\mathcal N}$ through~\cite{Guo:2008si}
\begin{eqnarray}
\sigma_n^{\rm SI} & = &{} \frac{1}{A^2}\frac{M^2(n)}{M^2({\mathcal N})}\sigma_{\mathcal N},
\end{eqnarray}
where $M(x) = m_DM_x/(m_D + M_x)$ with $M_{\mathcal N}$ being the target nucleus mass and $M_n = m_{p,n}$ being the nucleon mass.

\begin{figure}[htbp]
\begin{center}
\includegraphics[scale=0.6]{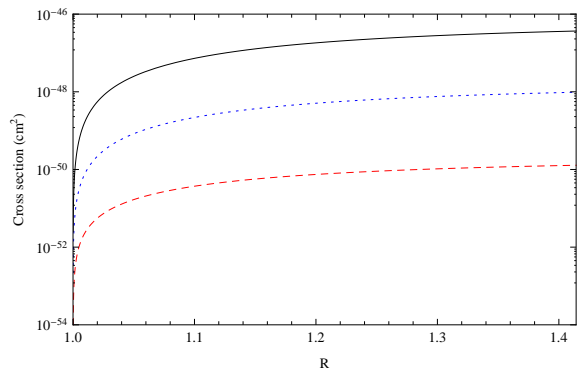}
\end{center}
\caption[]
{ The predicted spin-independent WIMP-nucleon elastic scattering cross section as a function of $R$ with $S = 1.0$ (solid line), $0.14$ (dotted line) and $0.01$ (dashed line). }
\label{fig:cross}
\end{figure}

In Fig.~\ref{fig:cross}, by taking the nuclei as Xenon with $Z= 54$ and  $A = 131.293$~\cite{Olive:2016xmw}, we plot the $R$ dependence of the normalized WIMP-nucleon elastic scattering cross section with typical choices of $S$. This figure shows that in the physically allowed parameter region, the cross section is much below the most stringent upper bound given by PandaX-II~\cite{Tan:2016zwf} so that it is reasonable to regarded it as a DM candidate. The results also tell us that the closer to the conformal region (the smaller $S$ and $R$), the smaller cross section.

We finally make some comments on the fermionic TC-baryon. The effective interaction between TC-baryon $\psi$ and electromagnetic field  is
\begin{eqnarray}
{\cal L}_{\psi}& = &{} - i eg_\psi \bar{\psi} \sigma^{\mu\nu}\partial_\alpha\psi F_{\mu\nu}.
\end{eqnarray}
where $g_\psi$ is the magnetic moment which can be calculated in the Skyrme approach. For isoscalar TC-baryon, one can obtain
\begin{eqnarray}
g_\psi & = &{} 3.2\times 10^{-3} \times \frac{e}{f_\pi}.
\end{eqnarray}
So that, from the geometric cross section one estimates $e \sim 10$ which indicates that the underlying TC theory should be far away from the conformal window. In this sense, we conclude that the number of technicolor cannot be odd.

\section{Summary and dicussions}

\label{sec:dis}

In this paper we explored the properties of the lightest EW baryon as a topological object in an effective Lagrangian of strongly coupled gauge theory of composite EWSB with respect to the present experimental constraint.

For the bosonic EW baryon which could be realized when the number of technicolor is even, we found that the small $S$ parameter, especially that with 95\% CL of the present constraint, which could be realized in SWC-TC induced a small EW baryon mass around a few hundred GeV. If we take into account the fact that near the conformal window the techni-$\rho$ and techni-$a_1$ are nearly degenerate, the EW baryon mass is further reduced to a few ten GeV. We conclude that in the composite EWSB theory, near the conformal window, in the case the chiral symmetry is restored which is a necessary condition for the conformal invariance, the small Peskin-Takeuchi $S$ parameter and the nearly degenerate techni-$\rho$ and techni-$a_1$ mass leads to a very light EW baryon with mass around a few ten GeV. In contrast, if the number of technicolor is odd, the EW baryon is a fermion. In this case, it has a unreasonable large mass so that is not favored phenomenologically.

We also calculated the cross section of the EW baryon-nuclei. We found that for a bosonic EW baryon, the cross is much below the most stringent PandaX-II constraint so that it is reasonable to regard the EW baryon as a DM candidate. However, for a fermionic EW baryon, to explain the relic density, the underlying fundamental theory should far away from the conformal window as a result the fermion EW baryon, or equally the odd number of technicolor, should be excluded.

In addition, since the bosonic EW baryon could have a mass at $O(100)$~GeV, it might be interesting to search it at certain collider such as LHC at CERN. Because the charge neutral bosonic EW baryon without EW quantum number interacts with the ordinary matter very softly, it is very difficult to observe it. But the charged bosonic EW baryon with EW quantum numbers which can be obtained by an appropriate arrange the EW quantum numbers of its constituents such TC fermion should be interesting.

As we know that from QCD that the soliton mass is affected by the infinite tower of vector resonances and the more resonances are included, the lighter baryon masses and smaller baryon sizes~\cite{Sutcliffe:2011ig,Ma:2012kb}. So that, the present results of the bosonic baryon mass and size can be regarded as a upper bound. A possible way to study the effect of the infinite tower of vector resonances is to the holographic models of strongly coupled theory. This will be reported elsewhere.


\acknowledgments

\label{ACK}

We would like to thank M.~Harada, M.~Rho, K.~Yamawaki and Y.~F.~Zhou for the stimulated discussions we had with them. The work was supported in part by National Science
Foundation of China (NSFC) under Grant No. 11475071, 11547308 and the Seeds Funding of Jilin
University.

\appendix


\begin{thebibliography}{99}


\bibitem{Peskin:2012vu}
M.~E.~Peskin,
arXiv:1208.5152 [hep-ph].  

\bibitem{:2012gk}
G.~Aad {\it et al.}  [ATLAS Collaboration],
Phys.\ Lett.\ B {\bf 716}, 1 (2012). 


\bibitem{:2012gu}
S.~Chatrchyan {\it et al.}  [CMS Collaboration],
Phys.\ Lett.\ B {\bf 716}, 30 (2012). 


\bibitem{Hill:2002ap}
  C.~T.~Hill and E.~H.~Simmons,
  Phys.\ Rept.\  {\bf 381}, 235 (2003)  [Erratum-ibid.\  {\bf 390}, 553 (2004)]. 


\bibitem{Yamawaki:1985zg}
  K.~Yamawaki, M.~Bando and K.~-i.~Matumoto,
  Phys.\ Rev.\ Lett.\  {\bf 56}, 1335 (1986).  

\bibitem{Appelquist:1998xf}
  T.~Appelquist and F.~Sannino,
  Phys.\ Rev.\ D {\bf 59}, 067702 (1999). 


\bibitem{Appelquist:1991is}
  T.~Appelquist and G.~Triantaphyllou,
  Phys.\ Lett.\ B {\bf 278}, 345 (1992).  

\bibitem{Sundrum:1991rf}
  R.~Sundrum and S.~D.~H.~Hsu,
  Nucl.\ Phys.\ B {\bf 391}, 127 (1993). 




\bibitem{Gipson:1980ag}
  J.~M.~Gipson and H.~C.~Tze,
  Nucl.\ Phys.\ B {\bf 183}, 524 (1981).  


\bibitem{Gipson:1983cm}
  J.~M.~Gipson,
  Nucl.\ Phys.\ B {\bf 231}, 365 (1984).  

\bibitem{D'Hoker:1984jq}
  E.~D'Hoker and E.~Farhi,
  Nucl.\ Phys.\ B {\bf 241}, 109 (1984).  

\bibitem{Nussinov:1985xr}
  S.~Nussinov,
  Phys.\ Lett.\  {\bf 165B}, 55 (1985). 


\bibitem{Murayama:2009nj}
  H.~Murayama and J.~Shu,
  Phys.\ Lett.\ B {\bf 686}, 162 (2010).

\bibitem{Joseph:2009bq}
  A.~Joseph and S.~G.~Rajeev,
  Phys.\ Rev.\ D {\bf 80}, 074009 (2009) . 



\bibitem{Gillioz:2010mr}
  M.~Gillioz, A.~von Manteuffel, P.~Schwaller and D.~Wyler,
  JHEP {\bf 1103}, 048 (2011) . 

\bibitem{Gillioz:2011dj}
  M.~Gillioz,
  JHEP {\bf 1202}, 121 (2012)  Erratum: [JHEP {\bf 1303}, 123 (2013)]  .



\bibitem{Ellis:2012bz}
J.~Ellis and M.~Karliner,
Phys.\ Lett.\ B {\bf 713}, 233 (2012). 


\bibitem{Ellis:2012cs}
  J.~Ellis, M.~Karliner and M.~Praszalowicz,
  JHEP {\bf 1303}, 163 (2013). 


\bibitem{Kitano:2016ooc}
  R.~Kitano and M.~Kurachi,
  JHEP {\bf 1607}, 037 (2016) .


\bibitem{Matsuzaki:2016iyq}
  S.~Matsuzaki, H.~Ohki and K.~Yamawaki,
   arXiv:1608.03691 [hep-ph].  

\bibitem{Bagnasco:1993st}
  J.~Bagnasco, M.~Dine and S.~D.~Thomas,
  Phys.\ Lett.\ B {\bf 320}, 99 (1994). 

\bibitem{Skyrme:1961vq}
  T.~H.~R.~Skyrme,
  Proc.\ Roy.\ Soc.\ Lond.\  A {\bf 260} (1961) 127.

\bibitem{Ma:2016gdd}
  Y.~L.~Ma and M.~Rho,
  Sci.\ China Phys.\ Mech.\ Astron.\  {\bf 60}, no. 3, 032001 (2017) . 

\bibitem{BReditor}
\newblock \textit{The Multifaceted Skyrmions: Second Edition}
(World Scientific, Singapore, 2016) ed.  M.~Rho and I. Zahed.


\bibitem{Bando:1987ym}
  M.~Bando, T.~Fujiwara and K.~Yamawaki,
  Prog.\ Theor.\ Phys.\  {\bf 79}, 1140 (1988).  


\bibitem{Bando:1987br}
  M.~Bando, T.~Kugo and K.~Yamawaki,
  Phys.\ Rept.\  {\bf 164}, 217 (1988).  

\bibitem{Tan:2016zwf}
  A.~Tan {\it et al.} [PandaX-II Collaboration],
  Phys.\ Rev.\ Lett.\  {\bf 117}, no. 12, 121303 (2016). 



\bibitem{Ma:2004sc}
  Y.~-L.~Ma, Q.~Wang and Y.~-L.~Wu,
  Eur.\ Phys.\ J.\ C {\bf 39}, 201 (2005). 




\bibitem{Peskin:1990zt}
  M.~E.~Peskin and T.~Takeuchi,
  Phys.\ Rev.\ Lett.\  {\bf 65}, 964 (1990).  

\bibitem{Peskin:1991sw}
  M.~E.~Peskin and T.~Takeuchi,
  Phys.\ Rev.\ D {\bf 46}, 381 (1992).  


\bibitem{Adkins:1983ya}
  G.~S.~Adkins, C.~R.~Nappi and E.~Witten,
  Nucl.\ Phys.\ B {\bf 228}, 552 (1983). 



\bibitem{Olive:2016xmw}
  C.~Patrignani {\it et al.} [Particle Data Group],
  Chin.\ Phys.\ C {\bf 40}, no. 10, 100001 (2016). 


\bibitem{Harada:2003dc}
  M.~Harada, M.~Kurachi and K.~Yamawaki,
  Phys.\ Rev.\ D {\bf 68}, 076001 (2003). 


\bibitem{Harada:2005br}
  M.~Harada and C.~Sasaki,
  Phys.\ Rev.\ D {\bf 73}, 036001 (2006). 

\bibitem{Bertone:2004pz}
  G.~Bertone, D.~Hooper and J.~Silk,
  Phys.\ Rept.\  {\bf 405}, 279 (2005) . 


\bibitem{Guo:2008si}
  W.~L.~Guo, L.~M.~Wang, Y.~L.~Wu, Y.~F.~Zhou and C.~Zhuang,
  Phys.\ Rev.\ D {\bf 79}, 055015 (2009) . 


\bibitem{Sutcliffe:2011ig}
  P.~Sutcliffe,
  JHEP {\bf 1104}, 045 (2011). 


\bibitem{Ma:2012kb}
  Y.~-L.~Ma, Y.~Oh, G.~-S.~Yang, M.~Harada, H.~K.~Lee, B.~-Y.~Park and M.~Rho,
  Phys.\ Rev.\ D {\bf 86}, 074025 (2012). 





\end{thebibliography}
\end{document}